\documentclass[12pt]{article}

\usepackage{times,epsfig,soul,graphics,amssymb,latexsym,amsmath,setspace,fullpage,enumerate,placeins,float}
\usepackage{array,threeparttable,hyperref,url}
\usepackage[usenames]{color}
\usepackage{booktabs, comment}
\usepackage{multirow}
\usepackage{algorithmicx}
\usepackage[affil-it]{authblk}
\usepackage{graphicx}
\usepackage[colorinlistoftodos]{todonotes}
\usepackage{authblk}
\usepackage{courier}
\usepackage{amsmath}
\usepackage{listings}
\usepackage{minted}
\usepackage{comment}
\title{Estimand framework and intercurrent events handling for clinical trials with time-to-event outcomes}
\author{Yixin Fang\thanks{Address: 1 North Waukegan Rd, North Chicago, IL 60064; email: yixin.fang@abbvie.com}\ \ and Man Jin}
\affil{Data and Statistical Sciences, AbbVie Inc.}
\date{\today}
\date{}

\usepackage{natbib}
\usepackage{graphicx}
\usepackage{amssymb}
\usepackage{setspace}

\DeclareMathAlphabet\mathbfcal{OMS}{cmsy}{b}{n}

\def\boxit#1{\vbox{\hrule\hbox{\vrule\kern6pt\vbox{\kern6pt#1\kern6pt}\kern6pt\vrule}\hrule}}

\begin{document}

\doublespacing

\maketitle

\begin{abstract}

The ICH E9(R1) guideline presents a framework of estimand for clinical trials, proposes five strategies for handling intercurrent events (ICEs), and provides a comprehensive discussion and many real-life clinical examples for quantitative outcomes and categorical outcomes. However, in ICH E9(R1) the discussion is lacking for time-to-event (TTE) outcomes. In this paper, we discuss how to define estimands and how to handle ICEs for clinical trials with TTE outcomes. Specifically, we discuss six ICE handling strategies, including those five strategies proposed by ICH E9(R1) and a new strategy, the competing-risk strategy. Compared with ICH E9(R1), the novelty of this paper is three-fold: (1) the estimands are defined in terms of potential outcomes, (2) the methods can utilize time-dependent covariates straightforwardly, and (3) the efficient estimators are discussed accordingly.

\end{abstract}


\section{Introduction}

The ICH E9(R1) guideline presents a framework for clinical trials to align planning, design, conduct, analysis, and interpretation \citep{ich2020e9}. The three key steps in the framework are: estimand, estimator, and sensitivity analysis \citep{mallinckrodt2020estimands}. ICH E9(R1) highlights the importance of dealing with intercurrent events (ICEs), which are defined as: 
\begin{quote}
    ``Events occurring after treatment initiation that affect either the interpretation or the existence of the measurements associated with the clinical question of interest. It is necessary to address intercurrent events when describing the clinical question of interest in order to precisely define the treatment effect that is to be estimated."
\end{quote}

ICH E9(R1) proposes five strategies for dealing with ICEs in clinical trials with quantitative outcomes and categorical outcomes: treatment policy strategy, hypothetical strategy, composite variable strategy, while-on-treatment strategy, and principal stratum strategy. The guideline provides many real-life clinical examples for quantitative outcomes and categorical outcomes, but lacks examples for time-to-event (TTE) outcomes. 
 
In this paper, we consider how to handle ICEs in clinical trials with TTE outcomes. This is a challenging task due to at least three reasons. First, how to handle censoring is challenging. In the literature of survival analysis, many censoring mechanisms have been discussed; e.g., fixed/random censoring, type I/II censoring, independent/dependent censoring, conditionally independent censoring, (non)prognostic censoring, (non)ignorable censoring, and (non)informative censoring. To solve such confusions, we consider censoring as a special case of missing data---whose mechanisms are well-known as missing completely at random (MCAR), missing at random (MAR), and missing not at random (MNAR)---and categorize censoring mechanisms into censoring completely at random (CCAR), censoring at random (CAR), and censoring not at random (CNAR). 

Second, how to utilize time-dependent covariates is challenging. Both the TTE outocome and the censoring may depend on the time-dependent covariates. To overcome this challenge, in this paper, we consider the targeted learning approaches \citep{van2011targeted}. 

Third, how to conduct sensitivity analysis to evaluate the robustness of the results to the deviation of the underlying assumptions (e.g., when the CAR assumption is violated, we consider the CNAR assumption) is also challenging. In this paper, we can extend the multiple-imputation (MI) method, which has been applied extensively in longitudinal studies for missing data, to be applicable in clinical trials with TTE outcomes. 

Fourth, there are different kinds of events. To distinguish among them, we refer to the event in the definition of the TTE outcome as the primary event (PE). Besides the censoring event we just discussed, there are other types of ICEs, which could be terminal or non-terminal. Handling terminal ICEs is challenging, because the occurrence of one terminal event would prevent the occurrence of the PE. Thus, in the literature of survival analysis, such terminal ICEs are referred to as competing events \citep{kalbfleisch2011statistical, prentice1978analysis}. For example, in a study examining time to death attributable to lung cancer, death attributable to other causes is a competing event. Therefore, we propose to add the competing risk strategy to the toolbox, which consists of the five strategies proposed by ICH E9(R1) already. 



The remaining of the paper is organized as follows. In Section 2, we describe the data structure. In Section 3, we overview the estimand framework and the targeted learning approaches. In Section 4, we discuss six ICE handling strategies, the five strategies proposed in ICH E9(R1) plus the competing risk strategy.  We conclude the paper with a summary in Section 5.

\section{Data structure} 

Theoretically, the TTE outcome variable is a continuous variable that is subject to being censored. Practically, we can consider the TTE outcome as a discrete variable in units of days, weeks, or months, assuming that the clinical trial is monitoring the subjects at daily, weekly, or monthly basis. Let $t=0, 1, 2, \dots, K$ index the time in a given unit, where $t=0$ indicates the the time zero (e.g., the time when the treatment is initiated) and $t=K$ indicates the final follow-up time. We now describe the discretization method that has been used widely in the literature of survival analysis; see for example, \cite{stitelman2012general}, \cite{benkeser2018improved}, \cite{van2011targeted}, and \cite{fang2024causal}.   

Let $W_i$ be the vector of baseline characteristics for subject $i$ measured at $t=0$. Let $A_i$ be the treatment (say, $1$ for the investigative treatment and $0$ for the control treatment) assigned to subject $i$ at $t=0$. Let $C_i(t-1)$ be the censoring status at the follow-up $t$,  $t=1, \dots, K$; in particular, $C_i(K-1)$ is the censoring status for the final follow-up time $K$. Let $Y_i(t)$ be the outcome variable at $t$, where $Y_i(t)=0$ means the PE hasn't occurred at $t$ and $Y_i(t)$ means the PE has occurred at $t$, $t=1, \dots, K$. One convention is that if $Y_i(t)=1$ then $Y_i(t')=1$ for $t'\geq t$. The other convention is that if $C_i(t-1)=\mbox{`censored'}$ then $Y_i(t')=\texttt{NA}$ for $t'\geq t$. 

The above data structure is for any time-independent treatment variable. We can generalize the data structure for a time-dependent treatment variable, defined as $A_i(t)$, $t=0, \dots, K-1$, to handle settings where there are treatment switches, treatment discontinuations, rescue medications, and so on. For the time-dependent treatment, we also collect time-dependent covariates $L_i(t)$, $t=1, \dots, K-1$ besides the outcome variables. Therefore, the time order of these variables is 
\begin{align}
W, A(0), C(0), Y(1), L(1), A(1), C(1), Y(2), L(2), \dots, A(K-1), C(K-1), Y(K), \label{eq:data}
\end{align}
where $L(K)$ is not needed if collected.

\begin{figure}[t!]
\centering
\includegraphics[height=9cm]{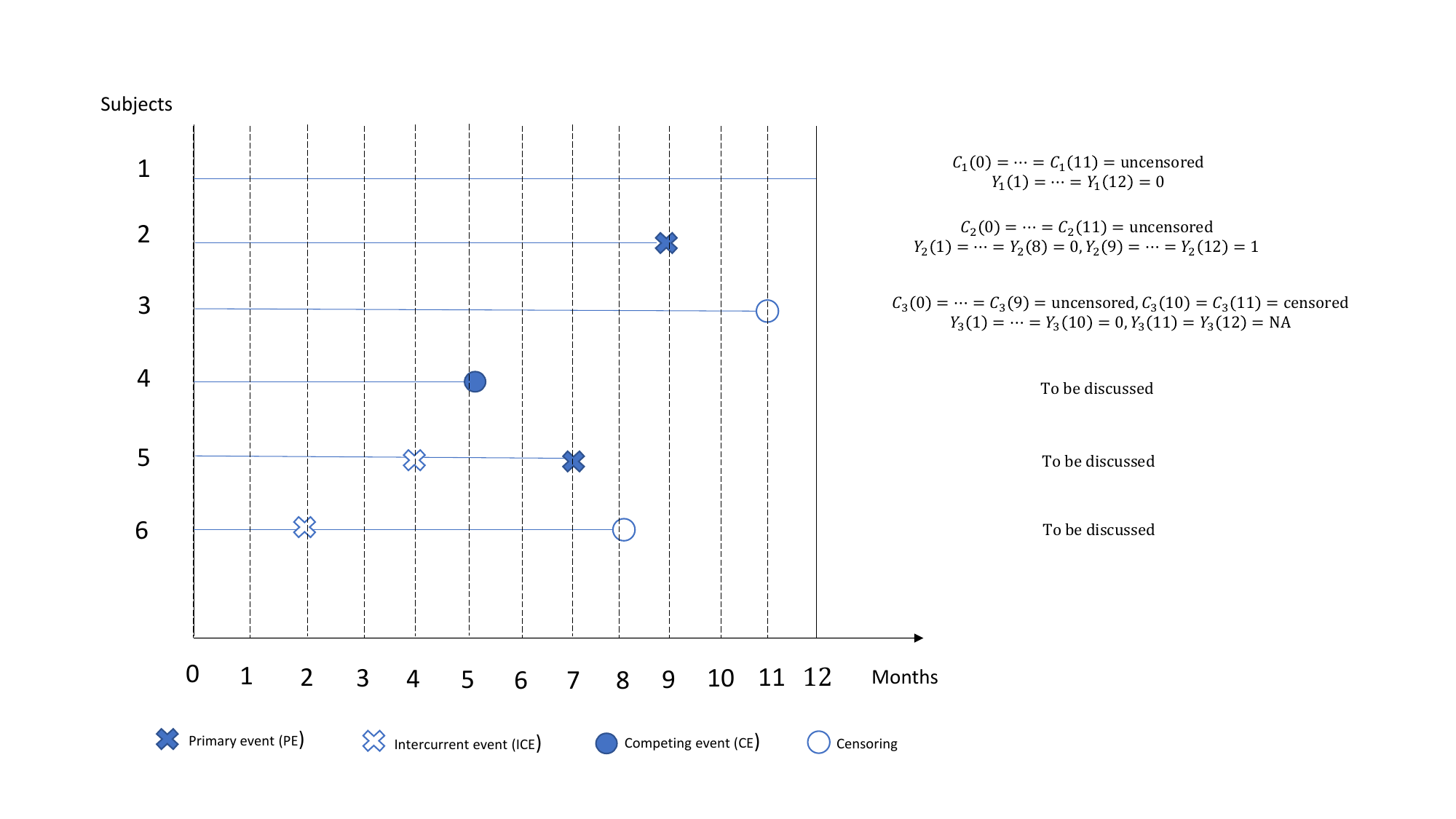}
\caption{Six hypothetical subjects}
\label{fig:figure-discretize}
\end{figure}

Figure \ref{fig:figure-discretize} shows six hypothetical subjects, assuming the PE is death from cancer. The 1st subject is alive at the end of the study (i.e., 12 months). Since there is no censoring, $C_1(0)=\dots=C_1(11)=\mbox{`uncensored'}$. Since the subject is alive throughout the study, $Y_1(1)=\dots=Y_1(12)=0$.  

The 2nd subject dies before the study ends. Since there is no censoring, $C_2(0)=\dots=C_2(11)=\mbox{`uncensored'}$. Since the subject dies at month 9, $Y_2(1)=\dots=Y_2(8)=0$ and $Y_2(9)=\dots=Y_2(12)=1$. 

The 3rd subject is censored before the study ends (e.g., loss to follow-up due to lack of efficacy). Since the subject is censored at month 11, $C_3(0)=\dots=C_3(9)=\mbox{`uncensored'}$ and $C_2(10)=C_2(11)=\mbox{`censored'}$. Since the subject is alive before month 10 while the status after censoring is unknown, $Y_3(1)=\dots=Y_3(10)=0$ and $Y_3(11)=Y_3(12)=\texttt{NA}$.

The 4th subject observes a competing event (e.g., death due to other causes) which precludes the occurrence of the PE. The competing event is a terminal ICE. We will discuss some strategies (e.g., the composite variable strategy or the competing risk strategy) for handling the competing event. Using different strategy will lead to different data of $C_4(t-1)$ and $Y_4(t)$, $t=1, \dots, 12$.

The 5th subject observes the PE before the study ends, but there is a non-terminal ICE (e.g., treatment discontinuation) that occurs before the PE occurs. The 6th subject is censored before the study ends, but there is a non-terminal ICE (e.g., rescue medication) that occurs before the subject is censored. Therefore, these two subjects are similar to the 2nd subject and the 3rd subject, respectively, except that there is a non-terminal ICE. For the 5th subject, if we ``ignore" the ICE, we have $C_5(0)=\dots=C_5(11)=\mbox{`uncensored'}$, $Y_5(1)=\dots=Y_5(6)=0$ and $Y_5(7)=\dots=Y_5(12)=1$. For the 6th subject, if we ``ignore" the ICE, we have  $C_6(0)=\dots=C_6(6)=\mbox{`uncensored'}$, $C_6(7)=C_6(11)=\mbox{`censored'}$, $Y_6(1)=\dots=Y_6(7)=0$, $Y_6(8)=Y_6(12)=\texttt{NA}$.

\section{Estimand framework}

\subsection{Estimand}

ICH E9(R1) defines estimand as:
\begin{quote}
 ``A precise description of the treatment effect reflecting the clinical question posed by the trial objective. It summarizes at a population-level what the outcomes would be in the same patients
under different treatment conditions being compared."
\end{quote}

Note that, in the above definition, ``the outcomes would be in the same patients under different treatment conditions being compared" is the so-called potential outcomes in the literature of causal inference \citep{imbens2015causal, fang2024causal}. 

Let $\overline{A}=(A(0), \dots, A(K-1))$ be the treatment sequence and let $\overline{a}=(a(0), \dots, a(K-1))$ be the treatment regime of interest. In particular, $\overline{a}=(1, \dots, 1)$ is the treatment regime of taking the investigative treatment throughout the study and $\overline{a}=(0, \dots, 0)$ is the treatment regime of taking the control treatment throughout the study. ICH E9(R1) proposes to define an estimand of interest according to its five attributes: population, treatment, outcome, ICE handling, and population-level summary. For example, we define an estimand of interest with the following five attributes,
\begin{itemize}
\item Population: to be defined by a set of inclusion/exclusion criteria,
\item Treatment variable: any feasible treatment regimes (either static or dynamic); e.g., a statistic treatment regime $\overline{a}$, 
\item Outcome variable: the time to the PE,
\item ICE handling: to be discussed in the next section in detail; while in this section, if censoring is considered as an ICE, it will be handled by the hypothetical strategy---envisaging a hypothetical scenario in which censoring would not occur---under the CAR assumption, 
\item Population-level summary: the survival function evaluated at a given time $t$, say $t=K$. 
\end{itemize}

Let $Y^{\overline{a}}(K)$ be the potential outcome at follow-up $K$ if the patient were treated by the treatment regime $\overline{a}$. Then the estimand of interest defined by the above five attributes is 
\begin{align}
    \theta=\mathbb{E}[1-Y^{\overline{a}}(K)],\label{eq:est}
\end{align}
which is the survival function at $K$ if all the patients in the population were treated by the treatment regime $\overline{a}$. Moreover, if we are interested in comparing two treatment regimes, $\overline{a}_1$ vs. $\overline{a}_0$, then the estimand of interest is defined as 
\begin{align}
    \theta=\mathbb{E}[1-Y^{\overline{a}_1}(K)]-\mathbb{E}[1-Y^{\overline{a}_0}(K)]=\mathbb{E}[Y^{\overline{a}_0}(K)]-\mathbb{E}[Y^{\overline{a}_1}(K)].\label{eq:est-01}
\end{align}

\subsection{Estimator}

The above estimands, (\ref{eq:est}) and (\ref{eq:est-01}), can be estimated using the targeted learning approaches \citep{van2006targeted, van2011targeted, van2018targeted}. In other words, the targeted learning approaches construct the targeted minimum loss-based estimators (TMLE), which enjoy many good statistical properties (e.g., consistency, normality, regularity, asymptotic linearity, efficiency, double robustness, substitution, and other semi-parametric properties) under the CAR assumption; see \cite{van2011targeted, van2018targeted} for detailed discussion of these properties. In this paper, we only discuss about the implementation using their R package ``ltmle" \citep{lendle2017ltmle}.

The R function ``ltmle" of the R package ``ltmle" has the following main arguments, ``Anodes", ``Cnodes", ``Lnodes", ``Ynodes", that are corresponding to the treatment variables, the censoring variables, the time-dependent variables, and the outcome variables in the data structure (\ref{eq:data}), 
\begin{align*}
    \mbox{Anodes} &= \{A(0), \dots, A(K-1)\},\\
    \mbox{Cnodes} &= \{C(0), \dots, C(K-1)\},\\
    \mbox{Lnodes} &= \{L(1), \dots, L(K-1)\},\\
    \mbox{Ynodes} &= \{Y(1), \dots, Y(K))\},
\end{align*}
along with the argument ``survivalOutcome=TRUE" indicating that the outcome is TTE (a.k.a., survival outcome). 

\subsection{Sensitivity analysis}

ICH E9(R1) defines sensitivity analysis as: 
\begin{quote}
  ``A series of analyses conducted with the intent to explore the robustness of inferences from the
main estimator to deviations from its underlying modeling assumptions and limitations in the
data."  
\end{quote}

The multiple imputation (MI) method is a popular method for handling missing data under the MAR assumption \citep{rubin2004multiple} and also a convenient method for conducting sensitivity analysis under the MNAR assumption \citep{o2014clinical, fang2022sequential}. Therefore, we can use MI for conducting sensitivity analysis under the CNAR assumption if the CAR assumption is violated. In the following three subsections, using a hypothetical dataset, we describe MI under the CAR assumption, under the copy-reference (CR) assumption, and under the jump to reference (J2R) assumption, respectively. Here CR and J2R are the two most popular reference-based methods for conducting sensitivity analysis \citep{o2014clinical}. 

\subsubsection{MI under CAR}

Consider a hypothetical RCT with two arms ($A=0, 1$) where the only ICE is censoring. Table \ref{tab:data:MI} shows 6 subjects, of whom 4 subjects are in the treated arm and 2 subjects are in the control arm. Although the baseline covariates and the time-dependent covariates are excluded from the table, they will be used in the imputation model.

Under the CAR assumption, we can apply the SAS procedure ``MI" or the R package ``mice" to perform MI. The original software is often only used for continuous or binary outcomes. But we can extend it easily to be applicable for TTE outcomes, as long as we follow the convention that if $Y_i(t)=1$ then $Y_i(t')=1$ for $t'\geq t$ to adjust the imputed outcomes.  

\begin{table}[t!]
\caption {A hypothetical dataset in which each arm has 200 subjects; e.g., subject 1 is censored at month 5, subject 2 dies at month 4, subject 400 is alive at the end of the study} \label{tab:data:MI} 
\begin{center}
\begin{tabular}{ cccccccccccccc} 
 \hline
ID & $A$ & FU1 & FU2 & FU3 & FU4 & FU5 & FU6 & FU7 & FU8 & FU9 & FU10 & FU11 & FU12\\
\hline
\multicolumn{14}{l}{Original dataset for MI under CAR}\\
\hline
1 & 0 & 0 & 0 & 0 & 0 & NA & NA & NA & NA & NA & NA & NA & NA \\
2 & 0 & 0 & $0$ & 0 & 1 & 1 & 1 & 1 & 1 & 1 & 1 & 1 & 1  \\
$\cdots$ & & & & & & & & & & & \\
397 & 1 & 0 & $0$ & 0 & 1 & 1 & 1 & 1 & 1 & 1 & 1 & 1 & 1 \\
398 & 1 & 0 & 0 &  0 & 0 & 0 & 0 & 0 & NA & NA & NA & NA & NA\\
399 & 1 & 0 & 0 & 0 & 0 & 0 & NA & NA & NA & NA & NA & NA & NA \\
400 & 1 & 0 & 0 & 0 & 0 & 0 & 0 & 0 & 0 & 0 & 0 & 0 & 0  \\
 \hline
 \hline
 \multicolumn{14}{l}{Tentative dataset for MI under CR}\\
\hline
1 & 0 & 0 & 0 & 0 & 0 & NA & NA & NA & NA & NA & NA & NA & NA \\
2 & 0 & 0 & $0$ & 0 & 1 & 1 & 1 & 1 & 1 & 1 & 1 & 1 & 1  \\
$\cdots$ & & & & & & & & & & & \\
398 & 0 & 0 & 0 &  0 & 0 & 0 & 0 & 0 & NA & NA & NA & NA & NA\\
399 & 0 & 0 & 0 & 0 & 0 & 0 & NA & NA & NA & NA & NA & NA & NA \\
\hline
\hline
 \multicolumn{14}{l}{Tentative dataset for MI under J2R}\\
\hline
1 & 0 & 0 & 0 & 0 & 0 & NA & NA & NA & NA & NA & NA & NA & NA \\
2 & 0 & 0 & $0$ & 0 & 1 & 1 & 1 & 1 & 1 & 1 & 1 & 1 & 1  \\
$\cdots$ & & & & & & & & & & & \\
398 & 0 & NA & NA &  NA & NA & NA & NA & NA & NA & NA & NA & NA & NA\\
399 & 0 & NA & NA & NA & NA & NA & NA & NA & NA & NA & NA & NA & NA \\
\hline
\end{tabular}
\end{center}
\end{table}

\subsubsection{MI under CR}

As described in \cite{fang2022sequential}, MI under the CR assumption consists of four steps: (1) create a tentative dataset, (2) apply MI under the MAR assumption, (3) adjust the dataset according the convention, and (4) recover the imputed dataset to the original form. 

Step 1: create a tentative dataset as demonstrated in Table \ref{tab:data:MI}, by deleting the complete cases in the treated arm and changing $A=1$ to $0$. After this, the tentative dataset has only one arm---the control arm, implying that the imputation based on this dataset will be under the CR assumption. 

Step 2: apply MI under the MAR assumption, utilizing the baseline covariates and time-dependent covariates, using either the SAS procedure ``MI" or the R package ``mice".

Step 3: adjust the imputed outcomes following the convention that if $Y_i(t)=1$ then $Y_i(t')=1$ for $t'\geq t$.

Step 4: recover the imputed dataset to the original form by changing back the arm variable to its original value and restoring the complete cases in the treated arm that are deleted in Step 1.

After the final dataset is obtained by the above 4 steps, we can apply the main estimation method (say, TMLE) to the final dataset. Repeat the process multiple times and combine the results using Rubin's rule. 

\subsubsection{MI under J2R}

As described in \cite{fang2022sequential}, MI under the J2R assumption also consists of four steps.

Step 1: create a tentative dataset as demonstrated in Table \ref{tab:data:MI}, by deleting the complete cases in the treated arm, changing the outcome variable at all the follow-ups in the treated arm to $\texttt{NA}$, and changing $A=1$ to $0$. After this, the tentative dataset has only one arm---the control arm---and all the values in the treated arm are tentatively censored after baseline, implying that the imputation based on this dataset will be under the J2R assumption. 

Step 2: apply MI under the MAR assumption, utilizing the baseline covariates and time-dependent covariates, using either the SAS procedure ``MI" or the R package ``mice".

Step 3: adjust the imputed outcomes following the convention that if $Y_i(t)=1$ then $Y_i(t')=1$ for $t'\geq t$.

Step 4: recover the imputed dataset to the original form by changing back the arm variable to its original value and restoring both the uncensored values and the complete cases in the treated arm that are deleted in Step 1.

After the final dataset is obtained by the above 4 steps, we can apply the main estimation method (say, TMLE) to the final dataset. Repeat the process multiple times and combine the results using Rubin's rule. 

\section{ICE handling strategies}

There are six ICE handling strategies, the five strategies proposed by ICH E9(R1) plus the newly added competing risk strategy. We describe them in the next six subsections respectively. 

We can categorize these ICE handling strategies into two categories, with Category One consisting of the composite variable, treatment policy, and hypothetical strategies and Category Two consisting of the remaining three strategies. 

\begin{figure}[t!]
\centering
\includegraphics[height=9cm]{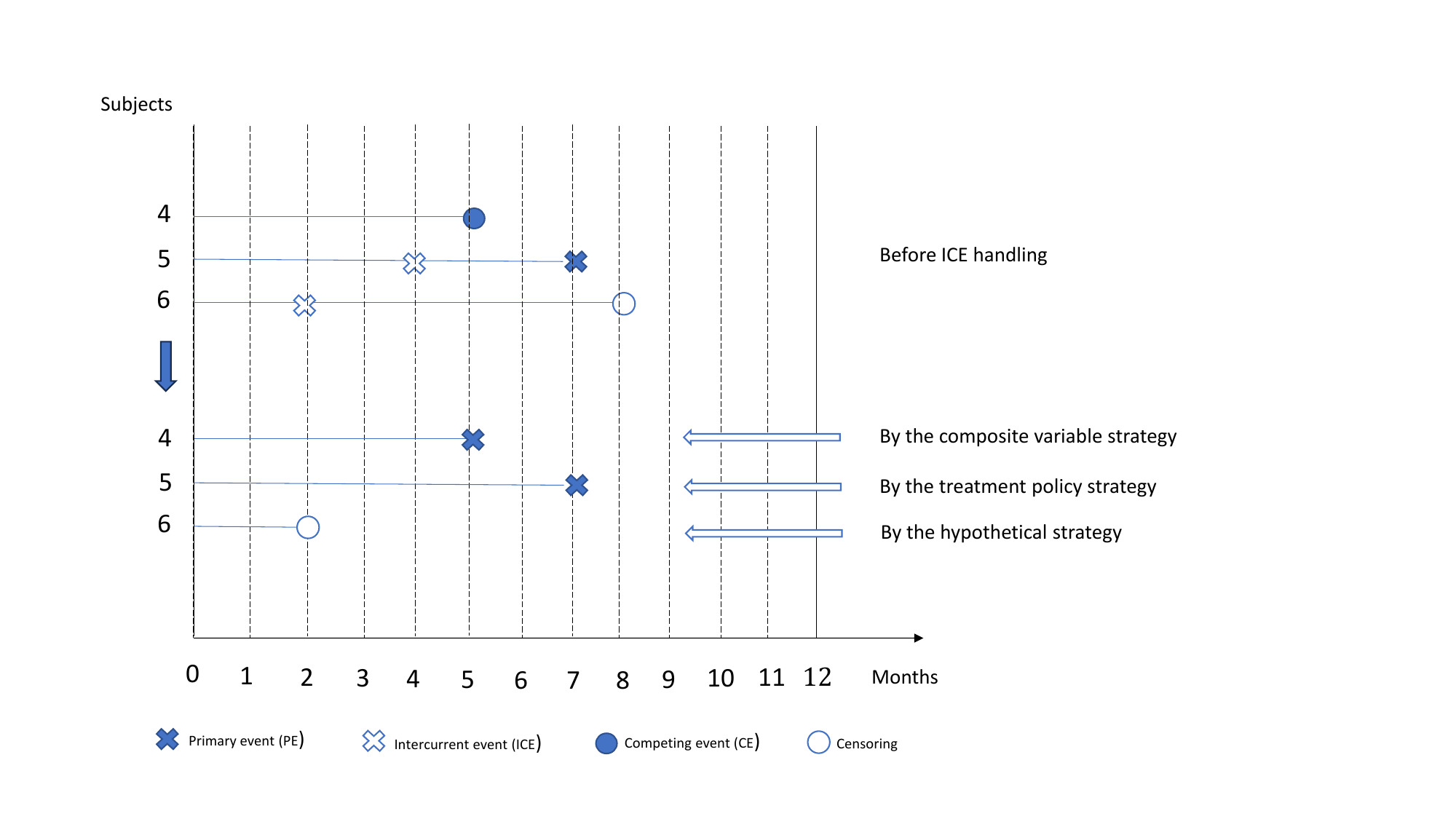}
\caption{Composite variable, treatment policy, and hypothetical strategies}
\label{fig:figure-strategy-one}
\end{figure}

The strategies in Category One can be demonstrated in Figure \ref{fig:figure-strategy-one}, in which we apply the three strategies in Category One to the 4th-6th hypothetical subjects in Figure \ref{fig:figure-discretize}. For the terminal ICE in the 4th subject (e.g., death due to other causes), we apply the composite variable strategy. For the non-terminal ICE in the 5th subject (e.g., treatment discontinuation), we apply the treatment policy strategy. For the non-terminal ICE in the 6th subject (e.g., rescue medication), we apply the hypothetical strategy.  

The other three strategies (the while-on-treatment, principal stratum strategy, and competing risk strategies), which form Category Two, are much more complex to understand than the ones in Category One, so in the paper we focus on the estimand definition and only describe the estimation method briefly. We will explore these estimation methods as our future work.

\subsection{The composite variable strategy}

\subsubsection{Estimand}

According to ICH E9(R1), by the composite variable strategy, ``an intercurrent event is considered in itself to be informative about the patient’s outcome and is therefore incorporated into the definition of the variable." By this strategy, we define a composite outcome variable, which is also an TTE outcome, that is, the time to either the PE or the ICE, whichever occurs first. 

The composite variable strategy can be applied to handle both terminal ICEs (including the competing event) and non-terminal ICEs. In summary, we can define an estimand of interest with the following five attributes: population, treatment, outcome (i.e., the newly defined TTE outcome), ICE handling (i.e., the composite variable for the ICE under consideration, and the hypothetical strategy for censoring), and population-level summary (i.e., the survival function). 

\subsubsection{Estimator}

We propose to estimate the above estimand using TMLE \citep{van2011targeted} with the corresponding R package ``ltmle" \citep{lendle2017ltmle}. To do so, we only need to update the definition of outcome variables in the ``Ynodes". For example, for the 4th subject in Figure \ref{fig:figure-strategy-one}, by the composite variable strategy, since the composite event (either the PE or the ICE under consideration) occurs at month 5, we have $Y_4(1)=\dots=Y_4(4)=0$, $Y_4(5)=\dots=Y_4(12)=1$, and  $C_4(0)=\dots=C_4(11)=\mbox{`uncensored'}$.

\subsection{The treatment policy strategy}

\subsubsection{Estimand}

According to ICH E9(R1), by the treatment-policy strategy, ``the occurrence of the intercurrent event is considered irrelevant in defining the treatment effect of interest: the value for the variable of interest is used regardless of whether or not the intercurrent event occurs." 

The treatment policy strategy can be applied to handle non-terminal ICEs (e.g., treatment discontinuation, treatment switching, and rescue medication). In summary, we can define an estimand of interest with the following five attributes: population, treatment, outcome (the time to the PE, or the time to the newly defined composite event if the composite variable strategy is applied to handle the terminal ICE), ICE handling (i.e., the treatment policy strategy for the ICE under consideration, the composite variable strategy for the terminal ICE if any, and the hypothetical strategy for censoring), and population-level summary (i.e., the survival function). 

\subsubsection{Estimator}

We propose to estimate the above estimand using TMLE \citep{van2011targeted} with the corresponding R package ``ltmle" \citep{lendle2017ltmle}. To do so, we only need to update the definition of treatment variables in the ``Anodes". 

For example, the two initial treatments under comparison are 0 and 1, then $A(t)=\texttt{NULL}$ means the subject discontinues the initially assigned treatment at $t$ and $A(t)=2$ means the subject takes the rescue medication at time $t$. For the 5th subject in Figure \ref{fig:figure-strategy-one}, we have $A_4(0)=A_4(1)=A_4(2)=A_4(3)$ and $A_4(4)=A_4(5)=A_4(6)=\texttt{NULL}$ or $2$ if the ICE occurring at month 4 is treatment discontinuation or rescue medication, respectively. By the treatment policy strategy, since the occurrence of the ICE is considered irrelevant, we adjust the Anodes such that $A_4(0)=\dots=A_4(6)$, along with having $C_5(0)=\dots=C_5(11)=\mbox{`uncensored'}$, $Y_5(1)=\dots=Y_5(6)=0$, and $Y_5(7)=\dots=Y_5(12)=1$. 

\subsection{The hypothetical strategy}

\subsubsection{Estimand}

According to ICH E9(R1), by the hypothetical strategy, ``a scenario is envisaged in which the intercurrent event would not occur: the value of the variable to reflect the clinical question of interest is the value which the variable would have taken in the hypothetical scenario defined." The hypothetical strategy requires that we use the counterfactual thinking \citep{pearl2009causality}. 

The hypothetical strategy can be applied to handle non-terminal ICEs (e.g., rescue medication). In summary, we can define an estimand of interest with the following five attributes: population, treatment, outcome (the time to the PE, or the time to the newly defined composite event if the composite variable strategy is applied to handle the terminal ICE), ICE handling (i.e., the hypothetical strategy for both the ICE under consideration and censoring), and population-level summary (i.e., the survival function). 

\subsubsection{Estimator}

We propose to estimate the above estimand using TMLE \citep{van2011targeted} with the corresponding R package ``ltmle" \citep{lendle2017ltmle}. To do so, we only need to update the definition of treatment variables in the ``Cnodes". 

For example, for the 6th subject in Figure \ref{fig:figure-strategy-one}, by the hypothetical strategy, we envisage a scenario in which the rescue medication would not be available and therefore the ICE under consideration would not occur." By this strategy, it is equivalent to censor the data after the occurrence of the ICE. Hence, for this subject, who is ``censored" at month 2, we have $C_6(0)=\mbox{`uncensored'}$, $C_6(1)=\cdots=C_6(11)=\mbox{`censored'}$, $Y_6(1)=0$ and $Y_6(2)=\cdots=Y_6(12)=\texttt{NA}$.

\subsection{The while-on-treatment strategy}

According to ICH E9(R1), by the while-on-treatment strategy, ``response to treatment prior to the occurrence of the intercurrent event is of interest. [...] If a variable is measured repeatedly, its values up to the time of the intercurrent event may be considered relevant for the clinical question." However, the TTE outcome cannot be measured immediately at the time of the occurrence of a non-terminable ICE, and a terminable ICE will preclude the occurrence of the primary event as well. Therefore, in ICH E9(R1), the while-on-treatment strategy is only discussed for quantitative and categorical outcomes, but not for TTE outcomes. 

Although the while-on-treatment strategy is not applicable to TTE outcomes, for completeness, we consider the following two alternatives. 

\subsubsection{Alternative one}

Consider a special case of the hypothetical strategy,  envisaging a hypothetical scenario in which, after the occurrence of a non-terminable ICE (e.g., treatment discontinuation, treatment switching, and rescue medication), the subject would not take any treatment. This is different from the hypothetical strategy discussed Subsection 4.2. The hypothetical strategy has several versions, among which these two versions---in the former version, the subject would continue the initially assigned treatment, while in this version, the subject would take no treatment after the occurrence of the ICE---are the most commonly used ones. 

Now assume that we apply both versions of the hypothetical strategy in the definition of an estimand of interest, assuming the former version and the new version of the hypothetical strategy are applied to ICE 1 and ICE 2, respectively. We can use a combination of the MI methods to estimate the estimand: (1) consider the data after the occurrence of ICE 1 or ICE 2 as censored, (2) use the MI under the CAR assumption to impute the censored data after ICE 1 while using the MI under the CR assumption to impute the censored data after ICE 2, and (3) apply the convention to adjust the imputed values. 

\subsubsection{Alternative two}

Instead of comparing the previous two treatment regimes, $(1, \dots, 1)$ versus $=(0, \dots, 0)$, we may be interested in comparing $\overline{a}_1=(1, \dots, 1, 0, \dots, 0)$---where the first $k$ components are ones the remaining components are zeros---versus $\overline{a}_0=(0, \dots, 0)$. The estimand of interest is defined as the one in (\ref{eq:est-01}) comparing the current $\overline{a}_1$ versus $\overline{a}_0$. 

We propose to estimate the above estimand using TMLE \citep{van2011targeted} with the corresponding R package ``ltmle" \citep{lendle2017ltmle}. To do so, we only need to specify the ``abar" argument in the R function ``ltmle"; i.e., \texttt{abar = list(treatment = c(rep(1, k), rep(0, K-k)), control = rep(0, K))}. 

\subsection{The competing risk strategy}

\subsubsection{Estimand}

To handle the ICE that is considered as a competing event (CE), we can use the composite variable strategy in Subsection 4.1 or the completing risk strategy to be discussed soon. In the literature of competing risk \citep{austin2016introduction}, a common population-level summary is the cumulative incidence function (CIF). 

To apply the competing risk strategy, we define a two-dimensional outcome variable $Y(t)=(Y_{PE}(t), Y_{CE}(t))$ under the convention that $Y_{PE}(t)$ and $Y_{CE}(t)$ cannot be equal to 1 simultaneously. That is, if $Y_{CE}(t)=1$, then $Y_{CE}(t')=1$ and $Y_{PE}(t')=0$ for all $t'\geq t$, and vice versa. 

Let $Y^{\overline{a}}_{PE}(K)$ be the potential outcome at follow-up $K$ if the patient were treated by the treatment regime $\overline{a}$. Then the estimand of interest is defined as 
\begin{align}
    \theta=\mathbb{E}[Y_{PE}^{\overline{a}}(K)],\label{eq:est-cr}
\end{align}
which is the CIF at $K$ if all the patients in the population were treated by treatment regime $\overline{a}$. 

\subsubsection{Estimator}

We propose to estimate the above estimand using TMLE \citep{van2011targeted} with the corresponding R package ``survtmle" (\underline{https://github.com/benkeser/survtmle}).

\subsection{The principal stratum strategy}

\subsubsection{Estimand}

Consider settings where the PE is non-terminal (e.g., hospital readmission) while the ICE is terminal (e.g., death) or settings where the PE is terminal (e.g., death) while the ICE is non-terminal (e.g., infection). The principal stratum strategy is applicable to handle the ICE in such settings. 

According to ICH E9(R1), by the principle-stratum strategy, ``the target population might be taken to be the principal stratum in which an intercurrent event would occur. Alternatively, the target population might be taken to be the principal stratum in which an intercurrent event would not occur. The clinical question of interest relates to the treatment effect only within the principal stratum." For settings where the PE is non-terminal while the ICE is terminal, we may be interested in the principal stratum in which the ICE would not occur. For settings where the PE is terminal while the ICE is non-terminal, we may be interested in the principal stratum in which the ICE would occur. Hereafter, we focus on the former settings. 

Let $Y_{D}(t)$ be the indicator of the occurrence of the terminal ICE under consideration at time $t$, $1\leq t\leq K$. According to ICH E9(R1), ``it is important to distinguish `principal stratification', which is based on potential intercurrent events [...] from subsetting based on actual intercurrent events." 
Let $Y^{1}_{D}(K)$ and let $Y^{0}_{D}(K)$ be potential intercurrent event occurrences at $K$ if the subjected were treated by $\overline{a}_1=(1, \dots, 1)$ and $\overline{a}_0=(0, \dots, 0)$, respectively. Assume that we are interested in the following principal stratum:
\begin{equation}
    PS_{AA} = \{i: Y^1_D(K)=0, Y^0_D(K)=0\},\label{eq:ps}
\end{equation}
which includes all the subjects who would be alive had been treated by either the two treatment regimes, where `AA' means being alive in both potential cases---note that there could be three other principal strata denoted by `AD', `DA', and `DD'. Then the estimand of interest can be defined as 
then the estimand of interest is defined as 
\begin{align}
    \theta=\mathbb{E}_{PS_{AA}}[1-Y_{PE}^{\overline{a}_1}(K)]-\mathbb{E}_{PS_{AA}}[1-Y_{PE}^{\overline{a}_0}(K)],\label{eq:est-01-PS}
\end{align}
where the expectation is taken over the principal stratum $PS_{AA}$ instead of the whole population. In the literature of survival analysis, the estimand is called the survivor average causal effect (SACE) \citep{robins1986new, zhang2003estimation}.

\subsubsection{Estimator}

In the literature of survival analysis, the problem of non-terminal outcomes that may be truncated by terminal ICEs is referred to as semi-competing risks because the terminal ICE acts as a competing risk for the non-terminal PE, but the reverse is not true \citep{fine2001semi}. However, the analyses of semi-competing risks emanating from the literature on survival analysis focus on estimating hazard ratios, without focusing on causal inference for treatment effects \citep{hernan2010hazards}. Recently, this gap was addressed by adapting the existing semi-competing risks methods and anchoring them to the principal-stratum framework for the purpose of causal inference \citep{comment2019survivor}.

\section{Summary}

ICH E9(R1) emphasizes the correct order for conducting clinical trials: ``having clarity in the trial objectives and accounting explicitly for intercurrent events when describing the treatment effect of interest at the planning stage should inform choices about trial design, data collection and statistical analysis." In this paper, we consider clinical trials with TTE outcomes and discuss six strategies for handling ICEs. We summarize the key findings in the following. 

The competing risk strategy can be applied to handle terminal ICEs along with terminal PE. The treatment policy strategy can be applied to handle non-terminal ICEs, if the data would be still collected after the ICE occurrence. The composite variable strategy can be applied to handle those ICEs that can be thought of as another mode of failure. The hypothetical strategy and the while-on-treatment strategy can be applied to those ICEs that are related to treatment compliance such as treatment discontinuation, treatment switching, and rescue medication. The principal stratum strategy can be applied to handle terminal ICEs along with non-terminal PE. 

When there are several types of ICEs, we can apply a combination of several strategies to handle them. For example, in a clinical trial, the PE is death due to lung cancer, and there are three types of ICEs: (1) treatment discontinuation and data would be collected after discontinuation, (2) death due to other causes, and (3) rescue medication. For this example, we may select the treatment policy for treatment discontinuation, the competing risk strategy for death due to other causes, and the hypothetical strategy for rescue medication.  

In the paper, we focus on the estimand definition and ICE handling, without providing a comprehensive review of the existing estimation methods. In this paper, we only mention a few estimation methods, such as TMLE, because it has nice asymptotical properties. In addition, we only touch a little bit on the topic of sensitivity analysis, which should be explored further.

\bibliographystyle{apalike}
\bibliography{references}

\begin{thebibliography}{}

\bibitem[Austin et~al., 2016]{austin2016introduction}
Austin, P.~C., Lee, D.~S., and Fine, J.~P. (2016).
\newblock Introduction to the analysis of survival data in the presence of competing risks.
\newblock {\em Circulation}, 133(6):601--609.

\bibitem[Benkeser et~al., 2018]{benkeser2018improved}
Benkeser, D., Carone, M., and Gilbert, P.~B. (2018).
\newblock Improved estimation of the cumulative incidence of rare outcomes.
\newblock {\em Statistics in medicine}, 37(2):280--293.

\bibitem[Comment et~al., 2019]{comment2019survivor}
Comment, L., Mealli, F., Haneuse, S., and Zigler, C. (2019).
\newblock Survivor average causal effects for continuous time: a principal stratification approach to causal inference with semicompeting risks.
\newblock {\em arXiv preprint arXiv:1902.09304}.

\bibitem[Fang, 2024]{fang2024causal}
Fang, Y. (2024).
\newblock {\em Causal Inference in Pharmaceutical Statistics}.
\newblock CRC Press.

\bibitem[Fang and Jin, 2022]{fang2022sequential}
Fang, Y. and Jin, M. (2022).
\newblock Sequential modeling for a class of reference-based imputation methods in clinical trials with quantitative or binary outcomes.
\newblock {\em Statistics in Medicine}, 41(8):1525--1540.

\bibitem[Fine et~al., 2001]{fine2001semi}
Fine, J.~P., Jiang, H., and Chappell, R. (2001).
\newblock On semi-competing risks data.
\newblock {\em Biometrika}, 88(4):907--919.

\bibitem[Hern{\'a}n, 2010]{hernan2010hazards}
Hern{\'a}n, M.~A. (2010).
\newblock The hazards of hazard ratios.
\newblock {\em Epidemiology (Cambridge, Mass.)}, 21(1):13.

\bibitem[ICH, 2020]{ich2020e9}
ICH (2020).
\newblock {ICH E9 (R1)} addendum on estimands and sensitivity analysis in clinical trials to the guideline on statistical principles for clinical trials.
\newblock {\em Accessed: 17th February}.

\bibitem[Imbens and Rubin, 2015]{imbens2015causal}
Imbens, G.~W. and Rubin, D.~B. (2015).
\newblock {\em Causal inference in statistics, social, and biomedical sciences}.
\newblock Cambridge University Press.

\bibitem[Kalbfleisch and Prentice, 2011]{kalbfleisch2011statistical}
Kalbfleisch, J.~D. and Prentice, R.~L. (2011).
\newblock {\em The statistical analysis of failure time data}, volume 360.
\newblock John Wiley \& Sons.

\bibitem[Lendle et~al., 2017]{lendle2017ltmle}
Lendle, S.~D., Schwab, J., Petersen, M.~L., and van~der Laan, M.~J. (2017).
\newblock ltmle: an r package implementing targeted minimum loss-based estimation for longitudinal data.
\newblock {\em Journal of Statistical Software}, 81:1--21.

\bibitem[Mallinckrodt et~al., 2020]{mallinckrodt2020estimands}
Mallinckrodt, C., Molenberghs, G., Lipkovich, I., and Ratitch, B. (2020).
\newblock {\em Estimands, Estimators and Sensitivity Analysis in Clinical Trials}.
\newblock CRC Press.

\bibitem[O'Kelly and Ratitch, 2014]{o2014clinical}
O'Kelly, M. and Ratitch, B. (2014).
\newblock {\em Clinical Trials with Missing Data: A Guide for Practitioners}.
\newblock John Wiley \& Sons.

\bibitem[Pearl, 2009]{pearl2009causality}
Pearl, J. (2009).
\newblock {\em Causality}.
\newblock Cambridge university press.

\bibitem[Prentice et~al., 1978]{prentice1978analysis}
Prentice, R.~L., Kalbfleisch, J.~D., Peterson~Jr, A.~V., Flournoy, N., Farewell, V.~T., and Breslow, N.~E. (1978).
\newblock The analysis of failure times in the presence of competing risks.
\newblock {\em Biometrics}, pages 541--554.

\bibitem[Robins, 1986]{robins1986new}
Robins, J. (1986).
\newblock A new approach to causal inference in mortality studies with a sustained exposure period—application to control of the healthy worker survivor effect.
\newblock {\em Mathematical modelling}, 7(9-12):1393--1512.

\bibitem[Rubin, 2004]{rubin2004multiple}
Rubin, D.~B. (2004).
\newblock {\em Multiple Imputation for Nonresponse in Surveys}, volume~81.
\newblock John Wiley \& Sons.

\bibitem[Stitelman et~al., 2012]{stitelman2012general}
Stitelman, O.~M., De~Gruttola, V., and Van~der Laan, M.~J. (2012).
\newblock A general implementation of tmle for longitudinal data applied to causal inference in survival analysis.
\newblock {\em The international journal of biostatistics}, 8(1).

\bibitem[van~der Laan and Rose, 2011]{van2011targeted}
van~der Laan, M.~J. and Rose, S. (2011).
\newblock {\em Targeted learning: causal inference for observational and experimental data}.
\newblock Springer Science \& Business Media.

\bibitem[van~der Laan and Rose, 2018]{van2018targeted}
van~der Laan, M.~J. and Rose, S. (2018).
\newblock {\em Targeted learning in data science}.
\newblock Springer.

\bibitem[van~der Laan and Rubin, 2006]{van2006targeted}
van~der Laan, M.~J. and Rubin, D. (2006).
\newblock Targeted maximum likelihood learning.
\newblock {\em The international journal of biostatistics}, 2(1).

\bibitem[Zhang and Rubin, 2003]{zhang2003estimation}
Zhang, J.~L. and Rubin, D.~B. (2003).
\newblock Estimation of causal effects via principal stratification when some outcomes are truncated by “death”.
\newblock {\em Journal of Educational and Behavioral Statistics}, 28(4):353--368.

\end{thebibliography}
\end{document}